\documentclass[aps,preprint]{revtex4}
\usepackage{amssymb}
\usepackage{amsmath}
\usepackage{graphicx}
\usepackage[normalem]{ulem}
\usepackage[dvips]{color}
\usepackage{pdfpages}

\renewcommand\sout{\bgroup \color{red} \ULdepth=-.5ex \ULset}

\renewcommand\sout{\bgroup \color{red} \ULdepth=-.5ex \ULset}

\begin{document}

\title{Noncommutativity and Holographic Entanglement Entropy}
\author{Tuo Jia\footnote{jiatuo23@physics.tamu.edu} and Zhaojie Xu\footnote{hanszee@tamu.edu}}
\affiliation{Department of Physics and Astronomy, Texas A\&M University,
College Station, TX 77843, USA}

\begin{abstract}
In this paper we study the holographic entanglement entropy in a large N noncommutative gauge field theory with two $\theta$ parameters by Ryu-Takayanagi prescription (RT-formula). We discuss what contributions the presence of noncommutativity will make to the entanglement entropy in two different circumstances: 1) a rectangular strip and 2) a cylinder. Since we want to investigate the entanglement entropy only, we will not be discussing the finite temperature case in which the entropy calculated by the area of minimal surface will largely be the thermal part rather than the entanglement part. We find that divergence of the holographic entanglement entropy will be worse in the presence of noncommutativity. In future study, we are going to explore the concrete way of computing holographic entanglement entropy in higher dimensional field theory and investigate more about the entanglement entropy in the presence of black holes/black branes.
\end{abstract}

\maketitle

\section{Introduction}
Over the past two decades, tremendous progress has been made in AdS/CFT duality and noncommutative geometry in string theory \cite{AH99,JMM99}. Noncommutativity has been widely explored in various cases since 40's in the last century \cite{HSS47,AC94,NS99,EW86,AC98,MRD01}. In noncommutative geoemtry and nonocommutative field theory, the coordinates of spacetime do not commute, which has many interesting properties. Meanwhile, the duality between supergravity in anti-de Sitter space and conformal field theory that lives on its boundary (AdS/CFT duality) has made great success \cite{JMM97,EW98,OA99}, and large changed our way of thinking about the nature. AdS/CFT has become a powerful tool for us to solve problems in quantum chromodynamics (QCD) \cite{TS04,TS05,SW15,SW16,TJ16}, condensed matter physics \cite{SAH09,CPH09,JM09,SS11}, etc. Noncommutative geometry shows up when we consider D-branes with B-field \cite{JMM99,NS99,CSS98,OB99,RB00}, and has many useful applications.

Entanglement entropy, being a fascinating topic as always, has been enhanced to a holographic version by Ryu and Takayanagi \cite{SR06}. The famous Ryu-Takayanagi formula was introduced as a formula of computing entanglement entropy of a system by computing the area of the minimal surfaces, which agrees the area law as in the calculation of the thermal entropy of black holes. However, a more concrete formula should be introduced when we deal with the entanglement entropy of higher dimensional field theory, we will leave this for future study.

What would be interesting is to consider combining holographic entanglement entropy and noncommutative field theory together. In \cite{WF13}, the holographic entanglement entropy of a large $N$-strongly coupled noncommutative field theory with one $\theta$ parameter has been well studied. In our work, we will consider a noncommutative gauge theory defined on $R_{\theta}^{2}\times R_{\theta}^{2}$, which has two $\theta$ parameters. This is the case when we turn on two components of B-field, we are going to compute the holographic entanglement entropy of the noncommutative gauge theory by Ryu-Takayanagi formula in two different circumstances: 1) a rectangular strip and 2) a cylinder. We will discuss what the additional noncommutativity contribute to the entanglement entropy by comparing our results with the results by Fishler, Kundu and Kundu \cite{WF13}. In this article, we will only discuss the spacetime without black holes, because when we introduce finite temperature by the presence of black holes, the entropy given by the RT-formula will not be the entanglement entropy only, the thermal entropy will also contribute to our results from the area of the minimal surfaces. Consequently, we will not discuss the finite temperature cases and we conjecture that the presence of black holes will not change the holographic entanglement entropy.

In section II, we will review the holographic entanglement entropy of a large $N$ strongly coupled noncommutative gauge field theory in an infinite rectangular strip when there is only one $\theta$ parameter, i.e., we turn on only one component of B-field. In section III, we turn on another component of B-field, and the resultant spacetime geometry will have an additional $\theta$ paremeter, the metric has an additional factor of $h(u)$. In order to compute the holographic entanglement entropy, we use RT-formula the compute the area of minimal surface. The entropy is simply given by
\begin{equation}
S_{RT}=\frac{Area(\gamma)}{4G},
\end{equation}
where $Area(\gamma)$ is the minimal value of an area functional and $G$ is Newton constant. We find that additional noncommutativity will make divergence of the holographic entanglement entropy worse in both cases: 1) an infinite noncommutative rectangular strip and 2) an infinite noncommutative cylinder. In section IV, we draw our conclusion and discuss some directions for future explorations.

\section{$\mathbb{R}^{2}\times \mathbb{R}_{\theta}^{2}$}
In this section, we review some results from \cite{WF13}, we study a large $N$-strongly coupled noncommutative field gauge theory with one $\theta$ parameter, i.e., on $R_{\theta}^{2}\times R^{2}$. The noncommutative parameter will only be on $R_{\theta}^{2}$ plane. Moyal algebra shows up in $R_{\theta}^{2}$, which is
\begin{equation}
[x^{2}, x^{3}]=i\theta.
\end{equation}
We can describe this noncommutative gauge field theory by holographic dictionary, and then we obtain the following metric in string frame \cite{AH99,JMM99}
\begin{equation}
ds^{2}=R^{2}\bigg[u^{2}f(u)dx_{0}^{2}+u^{2}dx_{1}^{2}+u^{2}h(u)(dx_{2}^{2}+dx_{3}^{2})+\frac{du^{2}}{f(u)u^{2}}+d\Omega_{5}^{2}\bigg],
\end{equation}
where
\begin{equation}
f(u)=1-\bigg(\frac{u_{H}}{u}\bigg)^{4},\qquad h(u)^{-1}=1+a^{4}u^{4}.
\end{equation}
Here $f(u)$ denotes the presence of a black hole. $h(u)$, which is caused by B-field, represents the existence of noncommutativity. 

On the other hand, we can compute the holographic entanglement entropy of a theory by RT-formula \cite{SR06} 
\begin{equation}
S_{RT}=\frac{Area(\gamma)}{4G},
\end{equation}
where $Area(\gamma)$ is the minimal value of an area functional, $G$ is the Newton constant. We can see that RT-formula has a lot in common with black hole entropy formula. Next, we will review the holographic entanglement entropy with the region being an infinite rectangular strip, we are going to omit the discussion of the noncommutative cylinder with one $\theta$ parameter and directly study the noncommutative cylinder with two $\theta$ parameters the compare our results with results from \cite{WF13}.

\subsection*{Noncommutative rectangular strip}
We consider an infinite rectangular strip, which is parameterized by
\begin{equation}
X=x^{2}\in[-\frac{l}{2},\frac{l}{2}],\qquad x^{1}, x^{3}\in [-\frac{L}{2},\frac{L}{2}]
\end{equation}
with $L\to\infty$. The area of the surface in the bulk is given by
\begin{equation}
\mathcal{A}=\frac{L^{2}R^{3}}{g_{s}^{2}}\int du u^{3}\sqrt{(X^{'2}+\frac{1}{h(u) f(u) u^{4}})}.
\end{equation}
The Lagrangian is a function of $X$, then the corresponding integrals of motion would give us
\begin{equation}
\frac{l}{2}=u_{c}^{3}\int_{u_{c}}^{u_{b}}\frac{du}{u^{5}\sqrt{(1-\frac{u_{c}^{6}}{u^{6}})hf}}.
\end{equation}
Plug the equation of constant motion back in the area functional, we get
\begin{equation}
\mathcal{A}=\frac{L^{2}R^{3}}{g_{s}^{2}}\int \frac{udu}{\sqrt{(1-\frac{u_{c}^{6}}{u^{6}})h(u)f(u)}}.
\end{equation}
\begin{figure}[h]
\centering
\includegraphics[scale=1]{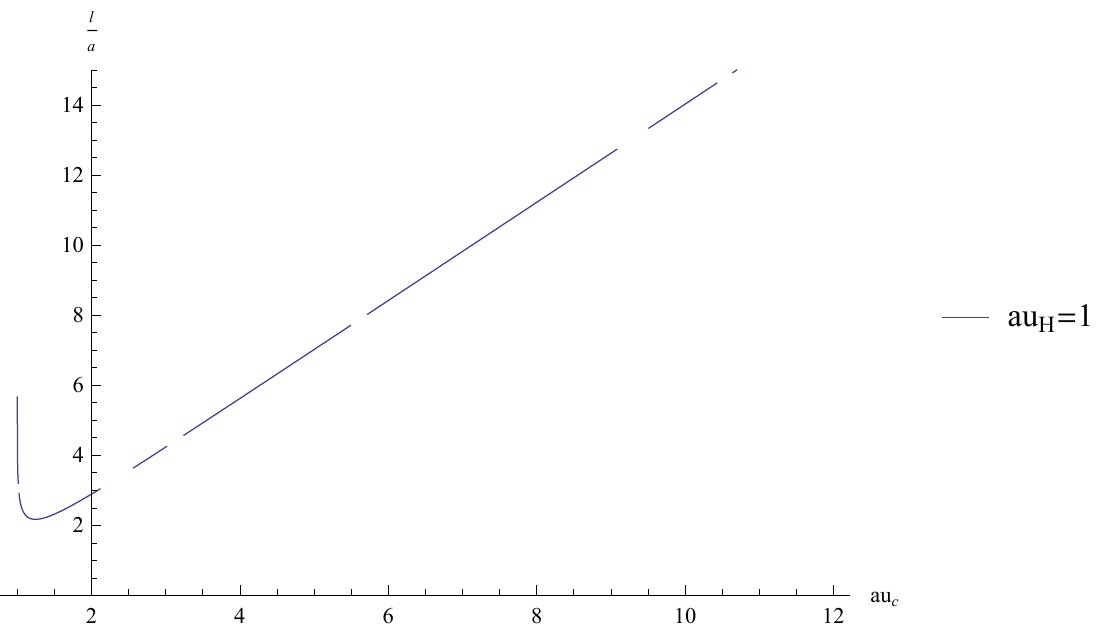}
\caption{Variation of l with $u_{c}$ for U shaped profile. We set $au_{H}=1$ and find that $l_{min}\sim 2.2a$ which is the result of the spacetime noncommutativity and finite temperature.}
\label{1}
\end{figure}

\section{$\mathbb{R}_{\theta}^{2}\times \mathbb{R}_{\theta}^{2}$}
In this section, we are going to turn on another component of B-field, i.e., now there are two $\theta$ parameters in the geometry and two functions $h_{1}(u)$ and $h_{2}(u)$ in the metric. The metric in the bulk is given by
\begin{equation}
ds^{2}=R^{2}\bigg[u^{2}h_{1}(u)(dx_{0}^{2}+dx_{1}^{2})+u^{2}h_{2}(u)(dx_{2}^{2}+dx_{3}^{2})+\frac{du^{2}}{u^{2}}\bigg],
\end{equation}
where
\begin{equation}
h_{i}^{-1}=1+a_{i}^{4}u^{4}.
\end{equation}
We make the convention $a_{1}=b, a_{2}=a$.

\subsection{Noncommutative rectangular strip}
The strip is parameterized by
\begin{equation}
X=x^{2}\in[-\frac{l}{2},\frac{l}{2}],\qquad x^{1}, x^{3}\in [-\frac{L}{2},\frac{L}{2}]
\end{equation}
with $L\to\infty$. The area of the surface in the bulk is given by
\begin{equation}
\mathcal{A}=\frac{L^{2}R^{3}}{g_{s}^{2}}\int du u^{3}\sqrt{(X^{'2}+\frac{1}{h_{2}u^{4}})\frac{1}{h_{1}}}.
\end{equation}
The constant of motion would give us
\begin{equation}
\frac{l}{2}=\frac{u_{c}^{3}}{h_{1}(u_{c})}\int_{u_{c}}^{u_{b}}\frac{du}{u^{5}}\sqrt{\frac{h_{1}(u)}{(1-\frac{u_{c}^{6}}{u^{6}})h_{2}(u)}},
\end{equation}
where $u_{c}$ represents the point closest to the extrmal surface.
\begin{figure}[h]
\centering
\includegraphics[scale=1]{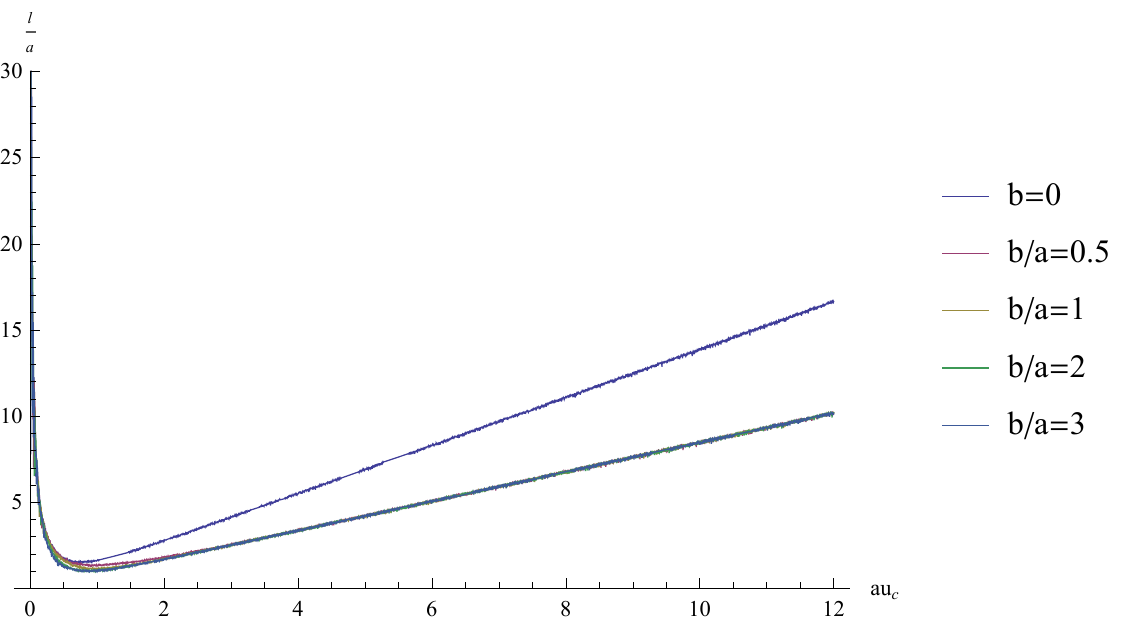}
\caption{Variation of l with $u_{c}$ for U shaped profile. For b=0 this reduce to the result of \cite{WF13}. For $b\neq 0$ the curves converge at large $u_{c}$ }
\label{1}
\end{figure}
Finally, we obtain the area functional, which is
\begin{equation}
\mathcal{A}=\frac{2L^{2}R^{3}}{g_{s}^{2}}\int_{u_{c}}^{u_{b}} du\,u\sqrt{\frac{1}{h_{1}(u)h_{2}(u)}\bigg(1+\frac{u_{c}^{6}h_{2}(u)}{(u^{6}-u_{c}^{6})h_{2}(u_{c})}\bigg)}.
\end{equation}
The divergence part of holographic entanglement entropy given by
\begin{equation}
\mathcal{A}_{div}=\frac{2L^{2}R^{3}}{g_{s}^{2}}\bigg[\frac{a^{2}b^{2}u_{b}^{6}}{6}+(\frac{a^{2}}{b^{2}}+\frac{b^{2}}{a^{2}})\frac{u_{b}^{2}}{4}\bigg].
\end{equation}
As a comparison, $\mathcal{A}_{div}$ with one $h(u)$ is 
\begin{equation}
\mathcal{A}_{div}^{'}=\frac{2L^{2}R^{3}}{g_{s}^{2}}\bigg[\frac{u_{b}^{2}}{4}+\frac{1}{2a^{4}}ln(au_{b})\bigg].
\end{equation}
We find that
\begin{equation}
\Delta \mathcal{A}_{div}\equiv\mathcal{A}_{div}-\mathcal{A}_{div}^{'}>0.
\end{equation}
Therefore, the divergence is even worsen by the additional noncommutativity.

\subsection{Noncommutative cylinder}
Now let's consider an infinite noncommutative cylinder
\begin{equation}
A=\{(x_{2},x_{3})|x_{2}^{2}+x^{2}_{3}\leq r^{2}\}, \qquad x^{1}\in[-L/2, L/2], \qquad L\rightarrow \infty,
\end{equation}
where $A$ is the circle that we construct in the $\{x^{2}, x^{3}\}$-plane.

We adopt the polar coordinates $dx_{1}^{2}+dx_{2}^{2}=d\rho^{2}+\rho^{2}d\theta^{2}$. The area functional is thus given by
\begin{equation}
\mathcal{A}=\frac{2\pi L R^{3}}{g_{s}}\int du u^{3}\rho(u)\sqrt{\bigg(\rho^{'}(u)^{2}+\frac{1}{u^{4}h_{2}(u)}\bigg)\frac{1}{h_{1}(u)}}.
\end{equation}
The equation of motion is
\begin{equation}
\frac{d}{du}\bigg(\frac{u^{3}\rho(u)\rho^{'}(u)}{\mathcal{L}_{0}}\bigg)=u^{3}\mathcal{L}_{0},
\end{equation}
where
\begin{equation}
\mathcal{L}_{0}=\sqrt{\bigg(\rho^{'}(u)^{2}+\frac{1}{u^{4}h_{2}(u)}\bigg)\frac{1}{h_{1}(u)}}.
\end{equation}

For arbitrary $h_{1}(u)$ and $h_{2}(u)$, the equation of motion is nearly impossible to solve. Therefore, we adopt the approximation $a^{4}u^{4}\gg1$ and $b^{4}u^{4}\ll1$. This seemingly unreasonable assumption that we make $b$ very small turns out to be well be reasonable, and it has something to do with electromagnetic duality where $a$ is the magnetic part and $b$ is the electric part. Electromagnetic duality (S-duality) requires that when one of them is very large, the other one has to be very small. However, it needs more concrete derivations, we leave it for future study.

In our approximation, we can compute the solution $\rho(u)$ in a series form. We are going to use the ansatz
\begin{equation}
\rho(u)=ku\bigg(1+\frac{c_{1}}{u^{4}}+\frac{c_{2}}{u^{8}}+\frac{c_{3}}{u^{12}}+\dots\bigg),
\end{equation}
where $k, c_{1}, c_{2}, c_{3}, \dots$ are constants. We find out that in the approximation $b^{4}u^{4}\ll1$, $b$ will not contribute to our series solution $\rho(u)$, the solution $\rho(u)$ is the same as the case which we only have one $B$ field! Therefore, the computation has been significantly simplified. The solution is
\begin{equation}
\rho(u)=\frac{a^{2}}{\sqrt{3}}u\bigg[1+\frac{1}{2a^{4}u^{4}}-\frac{1}{8a^{4}u^{4}}+\mathcal{O}\bigg(\frac{1}{a^{12}u^{12}}\bigg)\bigg],
\end{equation}
which dose not contain any contributions from $b$. In spite of that, we could try to solve the equation of motion with large $b$, we will find that the solution can not exist in any series form. The only way to solve the equation of motion is to assume $b$ to be very small, and this agrees with electromagnetic duality as well.

We now insert the solution $\rho(u)$ above to the area functional, we obtain the entanglement entropy
\begin{equation}
S(a, b)=N^{2}L\bigg[\frac{a^{4}b^{4}u_{b}^{9}}{27}+\frac{(2b^{4}/4+a^{4})u_{b}^{5}}{15}+\frac{1}{3}\bigg(1-\frac{b^{4}}{8a^{4}}\bigg)u_{b}\bigg].
\end{equation}
We compare it with the entropy when we only turn on one component of B field
\begin{equation}
S(a)=N^{2}L\bigg[\frac{2a^{4}u_{b}^{4}}{15}+\frac{u_{b}}{3}\bigg],
\end{equation}
we find out that the divergence of the entropy has only been worsen
\begin{equation}
\Delta S\equiv S(a, b)-S(a)=N^{2}Lb^{4}\bigg(\frac{a^{4}u_{b}^{9}}{27}+\frac{u_{b}^{5}}{30}-\frac{u_{b}}{24a^{4}}\bigg)>0.
\end{equation}
This indicates that when we turn on another component of B field, the entanglement entropy will only be more divergent.

As we can see from above, in both cases, $\Delta S>0$, which means additional noncommutativity will only worsen the divergence of the holographic entanglement entropy. This is a very interesting fact.

\section{CONCLUSION}
In this article, we first review the holographic entanglement entropy in a large $N$ strongly coupled noncommutative gauge field theory with only one $\theta$ parameter, i.e., we have only one $h(u)$ in the metric. Then we generalize the noncommutative field theory to the two $\theta$ parameter case. We have two $h(u)$'s in this case. By Ryu-Takayagagi formula, we compute the minimal surface for both regions: 1) an infinite noncommutative rectangular strip and 2) an infinite noncommutative cylinder. We find that in both cases, the divergence of the holographic entanglement entropy is worse in the presence of additional noncommutativity.

Another interesting fact in computing the holographic entanglement entropy in an infinite cylinder is that, we have assumed $b$ to be very small
\begin{equation}
b^{4}u^{4}\ll1, \qquad a^{4}u^{4}\gg1.
\end{equation}
This assumption agrees with electromagnetic duality with $a$ being the magnetic part and $b$ being the electric part. Electromagnetic duality (S-duality) makes this approximation natural and reasonable. However, the solid relation between electromagnetic duality and the holographic entanglement entropy in a noncommutative gauge field theory still needs more investigations in the future.

We haven't discussed the finite temperature noncommutative gauge field theory. The presence of a black hole will introduce a factor of $f(u)$ in the metric and surely will change the area of minimal surface, but the change of the entropy will largely be thermal part rather than entanglement part. We conjecture that the presence of black holes will not change the holographic entanglement entropy, it will contribute to the thermal entropy. For a detailed discussion, we will leave for future study.

Last but not least, we still need a more concrete formulation of computing (holographic) entanglement entropy in higher dimensional field theory. This requires fundamentally new ideas in both physics and mathematics. Despite that, we have shown that the additional noncommutativity will only worsen the divergence of the holographic entanglement entropy in a noncommutative field theory. We leave the above issues for future study.

\section*{ACKNOWLEDGMENTS}
We would like to thank Siu A. Chin, Si-wen Li and Yanli Shi for helpful discussions.

\end{document}